# First-principles investigation of mechanical properties of silicene, germanene and stanene

Bohayra Mortazavi[*,1], Obaidur Rahaman[r,1], Meysam Makaremi[2], Arezoo Dianat[3], Gianaurelio Cuniberti[3], Timon Rabczuk[1,#]

[1]*Institute of Structural Mechanics, Bauhaus-Universität Weimar, Marienstr. 15, D-99423 Weimar, Germany.*

[2]*Chemical Engineering Department, Carnegie Mellon University, Pittsburgh, Pennsylvania 15213, USA.*

[3]*Institute for Materials Science and Max Bergman Center of Biomaterials, TU Dresden, 01062 Dresden, Germany*

## Abstract

Two-dimensional allotropes of group-IV substrates including silicene, germanene and stanene have recently attracted considerable attention in nanodevice fabrication industry. These materials involving the buckled structure have been experimentally fabricated lately. In this study, first-principles density functional theory calculations were utilized to investigate the mechanical properties of single-layer and free-standing silicene, germanene and stanene. Uniaxial tensile and compressive simulations were carried out to probe and compare stress-strain properties; such as the Young's modulus, Poisson's ratio and ultimate strength. We evaluated the chirality effect on the mechanical response and bond structure of the 2D substrates. Our first-principles simulations suggest that in all studied samples application of uniaxial loading can alter the electronic nature of the buckled structures into the metallic character. Our investigation provides a general but also useful viewpoint with respect to the mechanical properties of silicene, germanene and stanene.

*Corresponding author (Bohayra Mortazavi): bohayra.mortazavi@gmail.com

Tel: +49 157 8037 8770,

Fax: +49 364 358 4511

[r] ramieor@gmail.com

[#]Timon.rabczuk@uni-weimar.de

## 1. Introduction

Graphene [1–3] is the planar form of sp[2] carbon atoms which is a semi-metal or zero-gap semiconductor. Graphene presents outstanding mechanical [4] and heat conduction [5] properties surpassing all known materials. The great success of graphene motivated a tremendous interest in the research for two-dimensional (2D) materials as a new class of materials with outstanding and tuneable properties. The wide application prospects of graphene has encouraged investments in the synthesis of other two-dimensional (2D) compounds including hexagonal boron-nitride [6,7], graphitic carbon nitride [8,9] silicene [10,11], germanene [12], stanene [13], transition metal dichalcogenides such as $MoS_2$ and $WS_2$ [14–16] and phosphorene [17,18]. One of the most attractive areas in the 2D materials research lies in their potential for integration. Various 2D materials can be integrated to form heterostructures [19,20] which not only provide a new class of materials with adjustable properties but also supply suitable building blocks for the next-generation electronic and energy conversion devices. A comprehensive understanding of the properties of 2D materials plays a crucial role in their real applications. Besides the advanced experimental investigations, theoretical studies can be considered as less expensive alternatives for the characterization of 2D material [21–26]. One of the key factors for the application of a material is its mechanical properties which correlate with its stability under the applied mechanical forces occurring during the operation.

The mechanical properties of silicene was previously studied by Qin et al. [27] using the first-principles method. They suggested that the in-plane stiffness of silicene is much smaller than that of graphene and the yielding strain of silicene under uniform expansion is about 20% in ideal conditions. On the other hand, Kaloni et al. [28] suggested that silicene lattice is stable up to 17% under biaxial tensile strain. In addition, first-principles calculations by Wang et al. [29] demonstrated a strain-induced self-doping phenomenon in both silicene and germanene nanosheets. They suggested that silicene and germanene have promising electronic properties that are absent in graphene and strain engineering can effectively tailor them toward applications in nanomaterials. Modarresi et al. [30] used density functional theory and molecular mechanics models to study the in plane-stiffness of stanene nanoribbons. They observed a closing of the energy gap in the band structure due to strain.

Although the elastic properties of silicene, germanene and stanene sheets have been studied in the past, little effort has been devoted to evaluate their stress-strain response and thus reporting the ultimate tensile strength and its corresponding strain. In addition, the chirality effect on the mechanical response has been less explored. In this work, we additionally present a direct comparison of these three nanomaterials under uniaxial strain. According to our results, these materials demonstrate similar mechanical responses, however, we qualitatively studied and compared several aspects like strain energy, stress-strain curves, evolution of bond lengths, buckling high and electronic density of states. We paid special attention to the chirality effect of these nanomaterials under strain. The knowledge about the mechanical response of these materials can provide very useful information for their applications in various systems such as those in nanoelectronics.

## 2. DFT Modeling

DFT calculations were performed as implemented in the Vienna ab initio simulation package (VASP) [31,32] using the Perdew-Burke-Ernzerhof (PBE) generalized gradient approximation exchange-correlation functional [33]. The projector augmented wave method [34] was employed with an energy cutoff of 450 eV. For all of the studied samples a super cell consisting of 64 atoms was fully relaxed with geometry optimization by using conjugate gradient method. Periodic boundary conditions were applied in all directions and a vacuum layer of 17 Å was considered to avoid image-image interactions along the sheet thickness. For the evaluation of mechanical properties the Brillouin zone was sampled using a 6×6×1 k-point mesh size and for the calculation of electronic density of states a single point calculation was carried out in which the Brillouin zone was sampled by employing a 15×15×1 k-point mesh size with the Monkhorst-Pack mesh [35]. After obtaining the optimized structure, uniaxial loading conditions were applied. To evaluate the mechanical properties using the uniaxial tensile simulations, we elongated the periodic simulation box size along the loading direction in multiple steps with a small engineering strain steps of 0.003. The applied elongation at the every step of loading, $\Delta L$, can be obtained based on the initial length of the unstrained sample along the loading direction, $L_0$, and the loading engineering strain which is equivalent to $\Delta L/L_0$. In a same manner, for the simulation of compression loading, we decreased the simulation box size along the loading direction using the calculated $\Delta L$. Then, in order to guarantee uniaxial stress condition in the sample, the simulation box size along the perpendicular direction of the loading was changed consequently such that the stress remained negligible in the perpendicular direction [23]. We note that by changing the simulation box size dimensions, we rescaled the atomic positions accordingly, so that no void is formed in the atomic lattice [23,26]. After applying the loading conditions, structural relaxation was achieved using the conjugate gradient energy minimization method with $10^{-5}$ eV criteria for energy convergence. The stress values at each loading strain step were then calculated to finally report the stress-strain relations for the considered structures.

## 3. Results and discussions

Uniaxial deformations of silicene, germanene and stanene stretched along the armchair direction are illustrated in Fig. 1. We depicted the structures at four different strain levels including no strain ($\varepsilon = 0.0$), one third of the strain at the ultimate tensile strength called, $\varepsilon_{uts}$ ($\varepsilon = 1/3\ \varepsilon_{uts}$), two third of $\varepsilon_{uts}$ ($\varepsilon = 2/3\ \varepsilon_{uts}$), and finally at $\varepsilon_{uts}$ ($\varepsilon = \varepsilon_{uts}$). For all of the structures, we observed a uniform extension along the loading direction. The buckling high of sheets also gradually decreases by increasing the strain level. Moreover, Fig. 1 reveals that in all cases the periodic sheet size along the transverse direction decreases due to the increase of the strain levels along the loading direction. For small strain levels within the elastic regime, the strain along the traverse direction ($\varepsilon_t$) with respect to the loading strain ($\varepsilon_l$) is acceptably constant. In this case one can evaluate the Poisson's ratio by calculating the ration of $-\varepsilon_t/\varepsilon_l$.

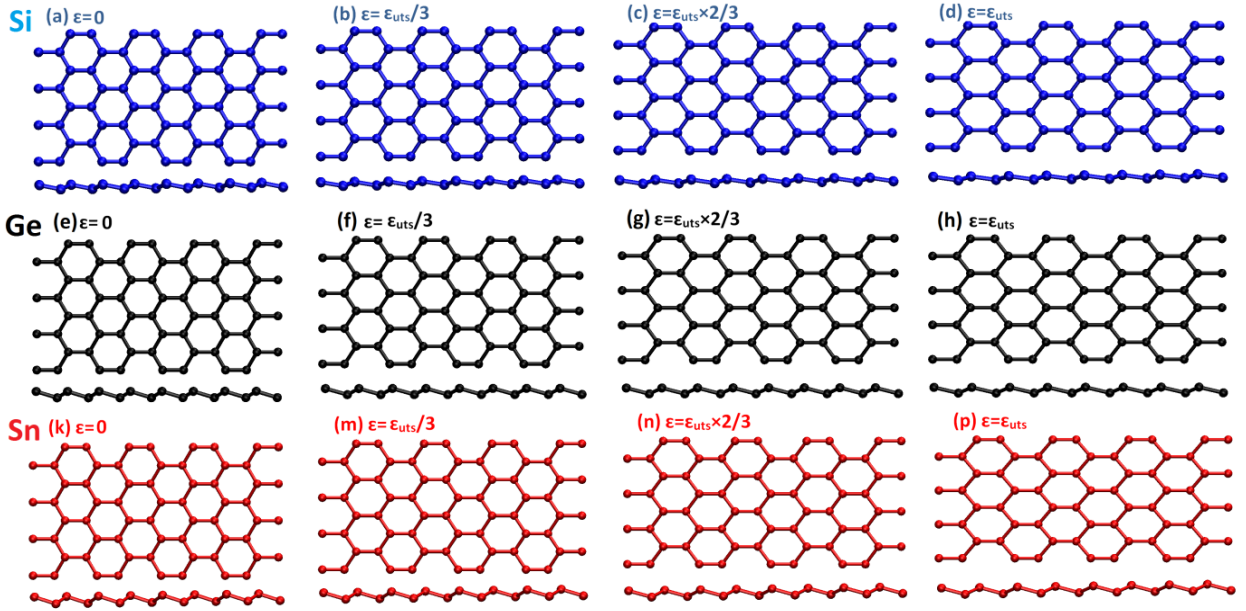

**Fig-1**, Top and side view of deformation processes of single-layer silicene (Si), germanene (Ge) and stanene (Sn) for different strain levels ($\varepsilon$) with respect to the strain at ultimate tensile strength ($\varepsilon_{uts}$). VMD [36] software is used to illustrate these structures.

In Fig. 2, strain energies of silicene, germanene and stanene under uniaxial compressive (negative strains) and tensile (positive strains) loading conditions are illustrated. To assess the effect of the loading direction, we performed uniaxial simulations along armchair and zigzag directions. The energy curves follow parabolic functions. By fitting a parabola (E = 0.5αe² + βe+c) to each energy curve, the elastic modulus can be calculated by $Y = \frac{1}{A}\alpha$. Here, A denotes the surface area of the sheet.

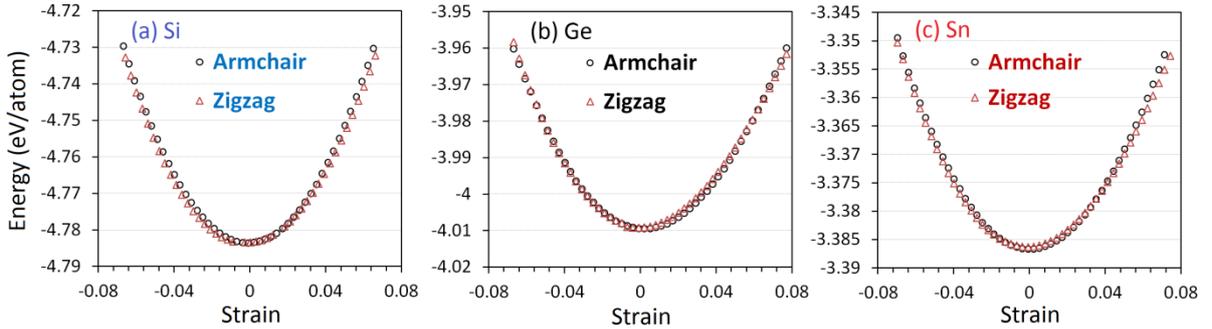

**Fig-2**, Strain energy of single-layer silicene (Si), germanene (Ge) and stanene (Sn) under uniaxial loading along armchair and zigzag.

**Table 1**, Mechanical properties of silicene, germanene and stanene sheets predicted by DFT method. Here, Y, P, STS and UTS depict elastic modulus, Poisson's ratio, strain at ultimate tensile strength point and ultimate tensile strength, respectively. Stress units are in GPa.nm.

| Structure | $Y_{armchair}$ | $Y_{zigzag}$ | $P_{armchair}$ | $P_{zigzag}$ | $STS_{armchair}$ | $STS_{zigzag}$ | $UTS_{armchair}$ | $UTS_{zigzag}$ |
|---|---|---|---|---|---|---|---|---|
| Silicene | 61.7 | 59 | 0.29 | 0.33 | 0.175 | 0.19 | 7.2 | 6.0 |
| Germanene | 44 | 43.4 | 0.29 | 0.35 | 0.2 | 0.205 | 4.7 | 4.1 |
| Stanene | 25.2 | 23.5 | 0.36 | 0.42 | 0.17 | 0.18 | 2.6 | 2.2 |

Acquired uniaxial stress-strain responses of defect-free and single-layer silicene, germanene and stanene along armchair and zigzag loading directions are illustrated in Fig. 3. For all studied cases, at the beginning the stress-strain curve has a linear formation which is followed by a nonlinear trend up to the ultimate tensile strength point. At this point the material presents its maximum load bearing ability and then the stress decreases by increasing the strain level. The strain at ultimate tensile strength point is also an important parameter which explains how much the material can be stretched before missing its load bearing ability due to the structural changes stemming from the uniaxial deformation. Our obtained results in Fig. 3 reveal that for all three studied 2D films, the sheets along the armchair direction can show higher tensile strengths with respect to the loading. Nonetheless, the strain at ultimate tensile strength is found not to remarkably dependent on the loading direction.

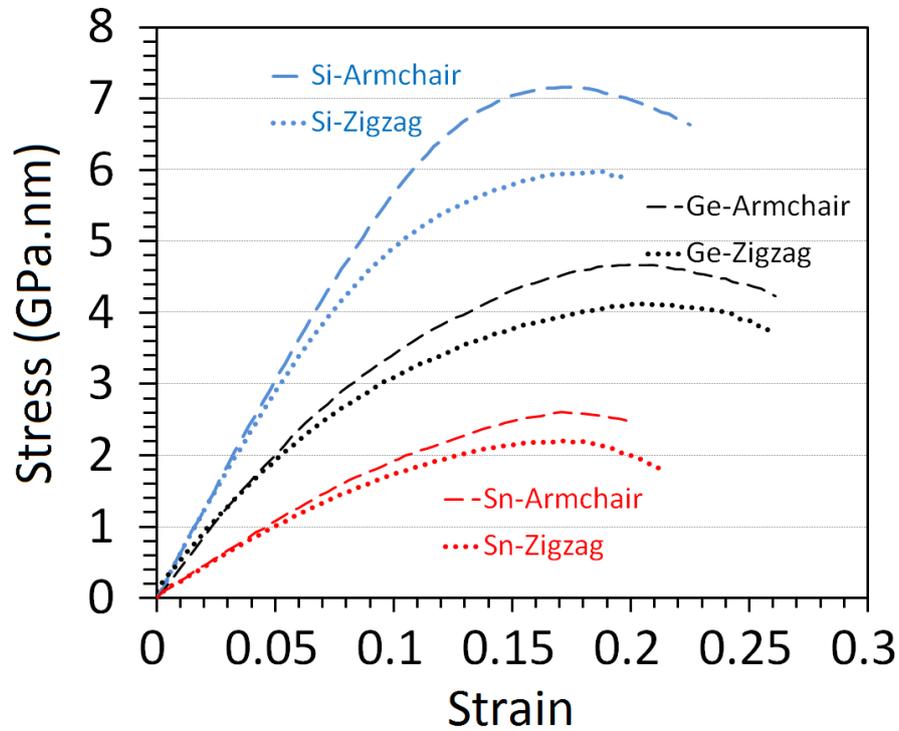

**Fig-3**, Calculated uniaxial tensile stress-strain response of defect-free and single-layer silicene (Si), germanene (Ge) and stanene (Sn) along armchair and zigzag loading directions.

**Table 2**, Comparison of elastic modulus of silicene, germanene and stanene sheets obtained in the present work with available information in the literature. The units are in GPa.nm.

|  | Elastic modulus (GPa.nm) | |
| --- | --- | --- |
|  | $Y_{armchair}$ | $Y_{zigzag}$ |
| **Silicene** | | |
| This work, DFT | 61.70 | 59.00 |
| Zhao [37], DFT | 63.51 | 60.06 |
| Qin et al. [27], DFT | 63.00 | 51.00 |
| John et al. [38], DFT | 61.33 | --- |
| **Germanene** | | |
| This work, DFT | 44 | 43.4 |
| John et al. [38], DFT | 42.05 | --- |
| **Stanene** | | |
| This work, DFT | 25.2 | 25.2 |
| John et al. [38], DFT | 24.46 | --- |
| Tao et al. [39], DFT | 24.14 | --- |

The mechanical properties of silicene, germanene and stanene sheets predicted by our DFT calculations are summarized in Table 1 and Table 2. As compared in Table 2, our calculated elastic modulus match well with previous theoretical predictions for all the three considered structures. The elastic modulus and ultimate tensile strength of the considered structures stretched along the armchair direction are generally higher than those stretched along the zigzag direction; whereas, the Poisson's ratio and the strain at the ultimate strength of the structures under the former tensile loading are lower than the ones under the latter loading. These behaviours of the considered 2D structures can be translated into higher strength in armchair direction, while more ductility in the zigzag direction.

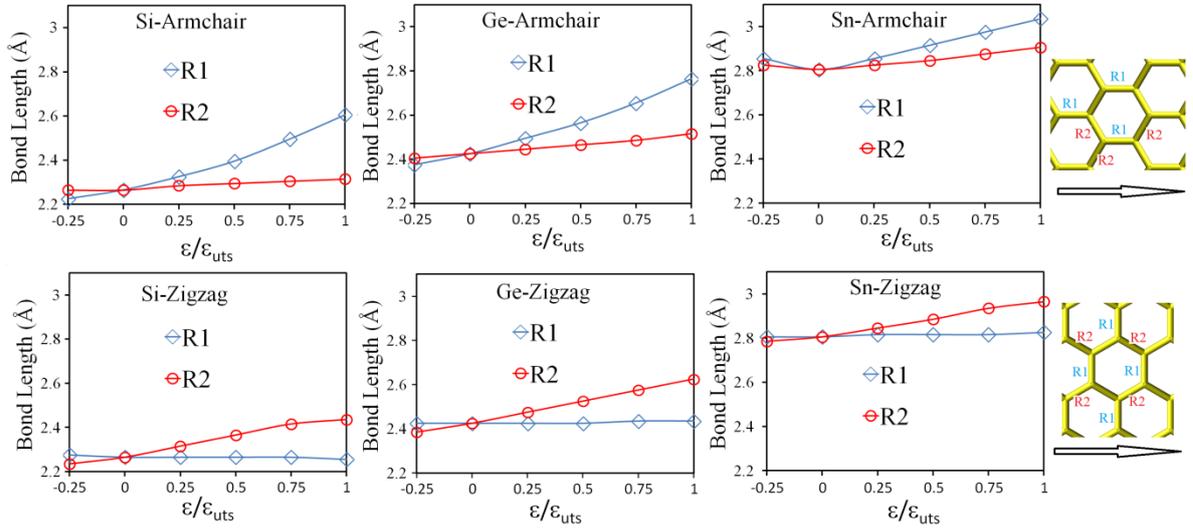

**Fig-4**, Evolution of bond lengths in silicene, germanene and stanene with increasing strain in the uniaxial direction. Results for stretching along both the armchair and zigzag directions are reported.

Fig. 4 shows the evolutions of bond lengths as the silicene, germanene and stanene structures are subjected to uniaxial loading. In most of the previous studies the bonds were not distinguished depending on their types and a monotonous increase in the bond length with increasing strain was observed [28]. However, this is less informative and interesting. In this work we distinguished the bonds depending on their orientations and analyzed their evolution with increasing strain. Stemming from the atomic radius, the bond length increases from silicene to stanene [40]. All of the bonds marked as R1 (and R2) had the same lengths during each step of uniaxial stretching for each particular structure, giving rise to a single peak in the radial distribution function. For all the cases, the peak positions for R1 and R2 are plotted in Fig. 4. It is worthwhile to note that stretching each structure along the armchair direction results in a slight but notable elongation of R2 in addition to the expected elongation of R1. On the other hand, when the structures were stretched along the zigzag direction, no notable change in R1 can be observed besides the expected elongation of R2. This is probably because of the fact that R2 bonds are oriented with an angle with respect to the armchair direction, thus stretching the structure along that direction can affect them, while R1 bonds are oriented perpendicular to the zigzag direction and may not be affected due to the tensile loading.

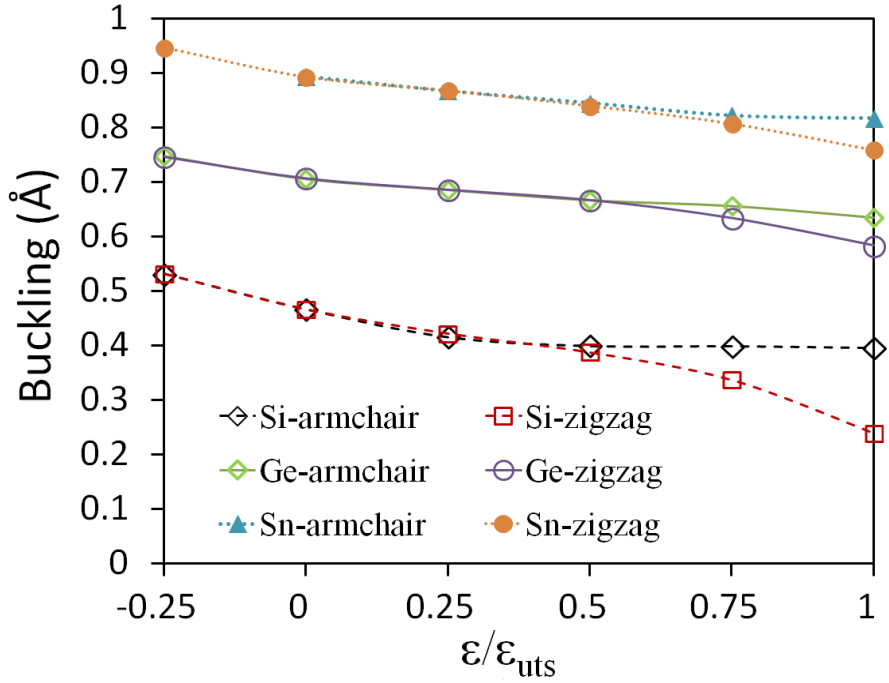

**Fig-5**, Buckling high of silicene, germanene and stanene structures as function of strain. Results for stretching along both the armchair and zigzag directions are reported.

Fig. 5 shows the buckling high of the silicene, germanene and stanene structures as function of the strain. At $\varepsilon/\varepsilon_{uts} = 0.0$, the bucking high for silicene, germanene and stanene includes 0.46 Å, 0.70 Å, 0.89 Å; respectively, which are in reasonable agreement with the data of previous studies [41,42] consisting of 0.45 Å, 0.69 Å, and 0.85 Å; accordingly. The data also suggests that stanene, germanene, and silicene respectively illustrate the highest to lowest buckling parameters. For all cases, the buckling high gradually reduces under the tensile loading. It is interesting to note that, for all of the three structures, the buckling high is similar irrespective of the direction of stretching until about $\varepsilon/\varepsilon_{uts} = 0.5$. After this point, for all cases when the structure is stretched along the zigzag direction the buckling high drops more steeply compared to that of the armchair direction. A sharp drop in the buckling high, signifying the flattening of the structure, is only observed for the case of silicene stretched along the zigzag direction.

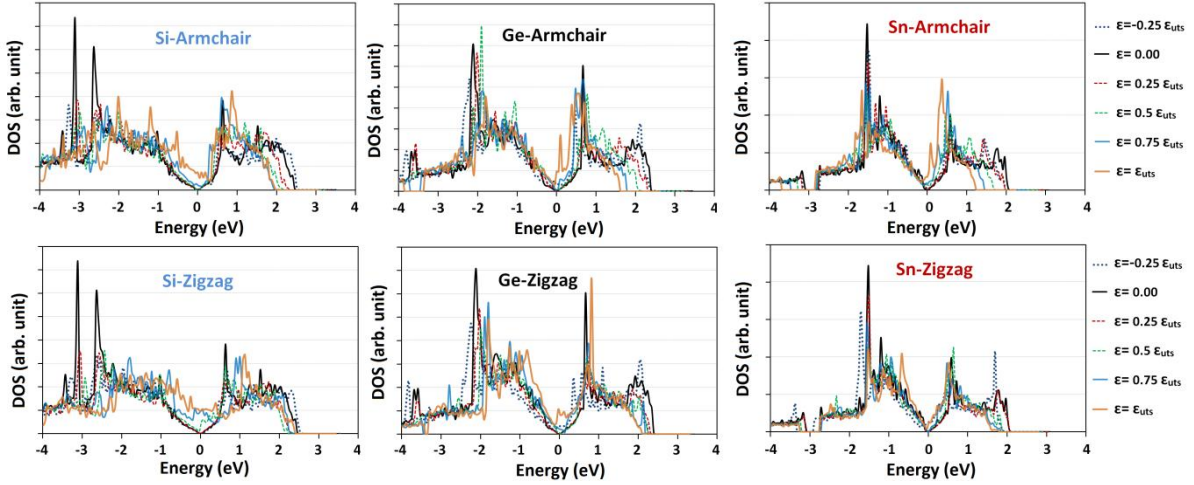

**Fig-6**, Electronic density of states (DOS) for silicene (Si), germanene (Ge) and stanene (Sn) for structure under different strains ($\varepsilon$) along the armchair and zigzag directions with respect to the strain at ultimate tensile strength ($\varepsilon_{uts}$).

Fig. 6 illustrates the calculated electronic density of states (DOS) for silicene, germanene and stanene for structures under different strains along the armchair and zigzag directions. In all cases, the relaxed film initially illustrates semiconductor properties with the bandgap of zero, well documented in the literature [40,43]; whereas, by applying the uniaxial compressive or tensile loading it is worth noting that the film show the metallic response which is the most pronounced at $\varepsilon/\varepsilon_{uts} = 1.0$. A recent computational study by Modarresi et al. [30] about mechanical properties of stanene confirms this phenomenon. A first-principles study conducted by Qin et al. [27] also suggested that the semimetal state of silicene can persist up to a tensile strain of 7%. Beyond that, silicene transforms into a conventional metal.

## 4. Summary

Mechanical characteristics of emerging 2D nanomaterials including single-layer silicene, germanene and stanene structures were investigated by performing DFT-PBE simulations. We used tensile loading simulations to study the effect of the chirality and the element of 2D materials on their mechanical properties. Silicene, germanene and stanene present the highest to the lowest tensile strengths, respectively. Loading along the armchair direction results in higher elastic modulus and tensile strengths; although, it leads to a lower Poisson's ratio and a smaller ultimate tensile elongation compared to the extension along the zigzag direction.

This work also predicts the armchair loading leads to the elongation of both bonds, whereas the zigzag loading only affects the length of the bond partially oriented along the loading direction and may not alter the bond lengths of the bonds perpendicular to the tension direction. In addition, For all structures the buckling high increases due to compressive loading, while it reduces under the tensile load. This effect is particularly pronounced for the extension along the zigzag direction at higher loading conditions. The electronic density of states calculations suggest that all of the 2D structures may illustrate different electronic properties with respect to the magnitude of the loading. At equilibrium and under no loading, they illustrate zero bandgap semiconducting properties, while at intense compressive or tensile loading they may show a perfect metallic behavior.

## Acknowledgment

BM, OR and TR greatly acknowledge the financial support by European Research Council for COMBAT project (Grant number 615132).